\documentstyle[pre,aps,multicol,epsf]{revtex}

\begin{document}   

%\draft				     
\tightenlines

\title{Comment on ``Phase ordering in chaotic map lattices with conserved
dynamics''}
\author{Julien Kockelkoren and Hugues Chat\'e}
\address{CEA --- Service de Physique de l'Etat Condens\'e, 
Centre d'Etudes de Saclay, 91191 Gif-sur-Yvette, France}
\maketitle
\begin{abstract}
Angelini, Pellicoro, and Stramaglia [Phys. Rev. E {\bf 60}, R5021 (1999)]
(APS) claim that the phase ordering of two-dimensional
systems of sequentially-updated
chaotic maps with conserved ``order parameter'' does not belong, for
large regions of parameter space, to the expected universality class.
We show here that these results are due to a slow crossover and that a
careful treatment of the data yields  normal dynamical 
scaling. Moreover, we construct better models, i.e. 
synchronously-updated coupled map lattices, which are exempt from these 
crossover effects, and allow for the first precise estimates of 
persistence exponents in this case.
\end{abstract}
%%%XXX put right PACS

\begin{multicols}{2} 

\narrowtext 

\section{Introduction}

The dynamical phenomenon of domain growth occurs in many different physical 
contexts. Once a fairly well established subject \cite{BRAY}, 
it has been recently
the center of renewed interest for several distinct reasons: one is
the advent of new quantifiers of the  associated dynamical scaling regimes,
such as first-passage or persistence exponents, i.e. the rate of algebraic
decay of the probability for a given point in space to have
remained in the same phase since some initial time \cite{PERSIS}.
Another reason is the natural question of the extent of known
``universality classes'' to new types of systems, e.g. spatio-temporally
chaotic ones.

In this context, the recent study of the ordering properties
of chaotic coupled map lattices (CMLs) possessing several symmetric phases
in competition brought up some intriguing results \cite{CMLPER}: 
in the simple case of
two competing phases and a non-conserved order parameter, the
``normal'' growth law $L(t)\sim t^{1/z}$ with $z=2$, 
where $L(t)$ is the single lengthscale characterizing the coarsening pattern,
was observed but with some exponent $z\ge 2$ continuously varying with 
parameters. However, this was later shown \cite{DDAP99} to be only a (slow)
transient behavior due to the non-trivial effect of space-discretization
in these deterministic systems. For larger lattices and longer times than
those considered in \cite{CMLPER}, $L$ was shown to behave normally
when plotted against $t^{1/2}$.

In a recent Rapid Communication, Angelini, Pellicoro, and Stramaglia (APS)
\cite{APS}, motivated by the above study of non-conserved order
parameter CMLs, presented a class of sequentially-updated lattices of
chaotic maps designed to investigate the case where the order parameter is 
locally conserved. In this case, $L$ is also expected to grow
algebraically with time, but with $z=3$ \cite{BRAY,HUSE}. 
APS claim, however, that
larger exponents are commonly found. Here we show  that APS were misled by 
their treatment of data and 
that in fact the
normal ($z=3$) growth law is observed in all cases. We argue, moreover, that
fully deterministic, synchronously-updated, coupled map lattices 
which conserve the order parameter can be easily constructed following the
ideas of Oono and Puri \cite{OONO-PURI}, 
and we show that these systems behave very smoothly,
enabling the precise measurement of persistence exponents in this context.

\section{Revisiting APS results}

\subsection{The hybrid map lattices of APS}

The lattices of maps introduced by APS are hybrid in several ways:
a given local map $f$ is first applied to all sites $x_i$ 
of the lattice (a deterministic
and synchronous operation), then pairs of nearest neighbors are sequentially
and regularly visited and swapped probabilistically. (The regularity of
the sweeps of the lattice is at the origin of the anisotropy of the 
domains in Fig.~1 of \cite{APS}.)
Furthermore, such systems are not ``coupled'' maps as in usual CMLs,
since the values taken by the sites are not influenced by the swaps 
(they are always taken according to the invariant measure of the local map).

These systems are designed to mimick Ising systems (with fluctuating couplings)
corresponding to the ``spins'' $\sigma_i = {\rm sgn}(x_i)$.
The local map $f$ has not much importance, and it is convenient to
choose, following \cite{MH,CMLPER}, an odd map of the $[-1,1]$ interval
with two symmetric attractors..
The energy of
one configuration is given by $E = -\sum_{\langle i,j \rangle} x_i
x_j$ where the sum is over nearest-neighbor pairs.
The exchange probability reads
$P_{\rm swap} = 1/(1+\exp(\beta\Delta E)$ where $\beta$ is the inverse
temperature and $\Delta E$ is the energy change of the swap.
The zero-temperature limit is deterministic: swaps are effective if and only if
they decrease the energy.

\subsection{Domain growth is normal}

\begin{figure}
\centerline{
\epsfxsize=4truecm
\epsffile{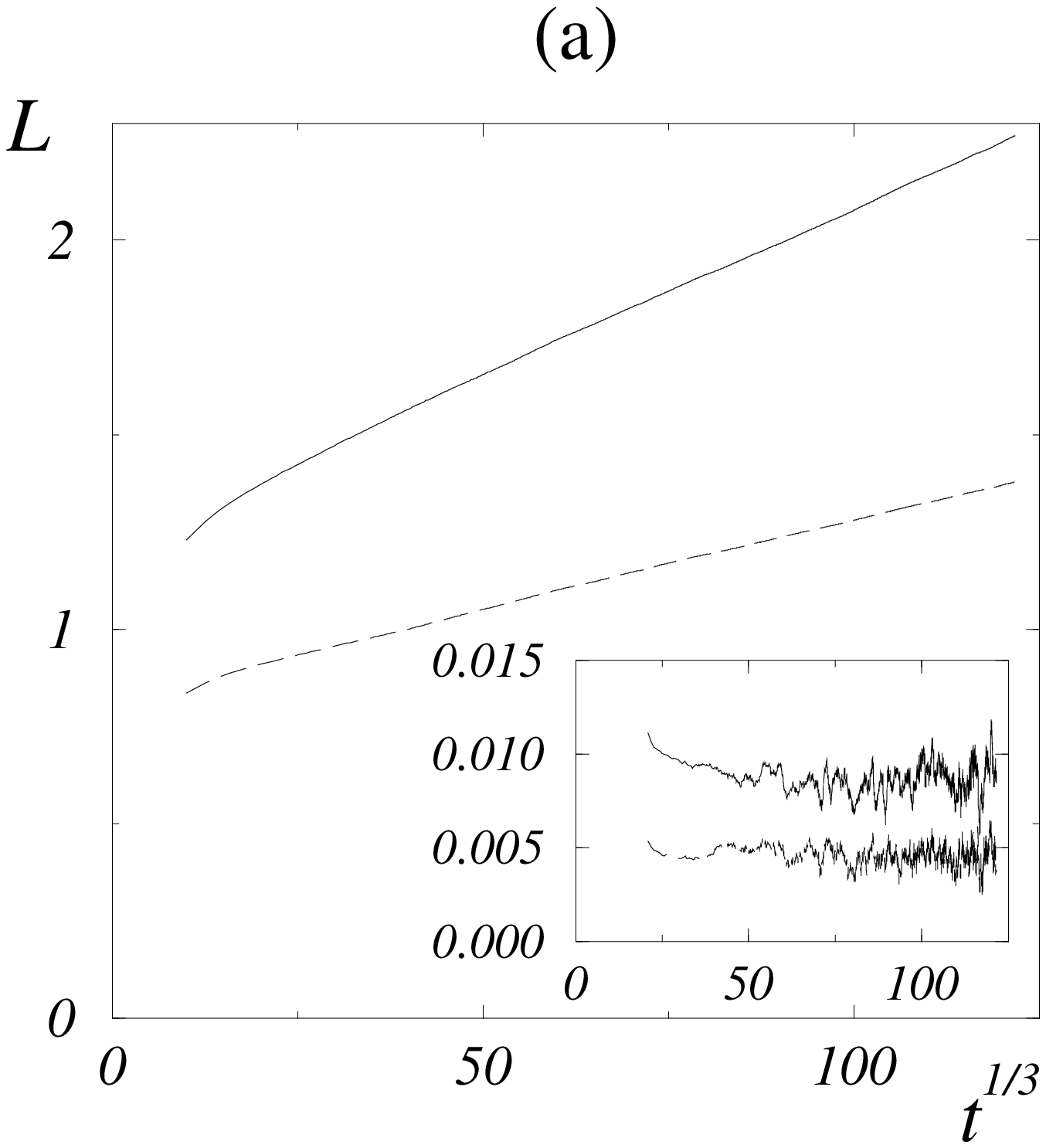}
\epsfxsize=4.5truecm
\epsffile{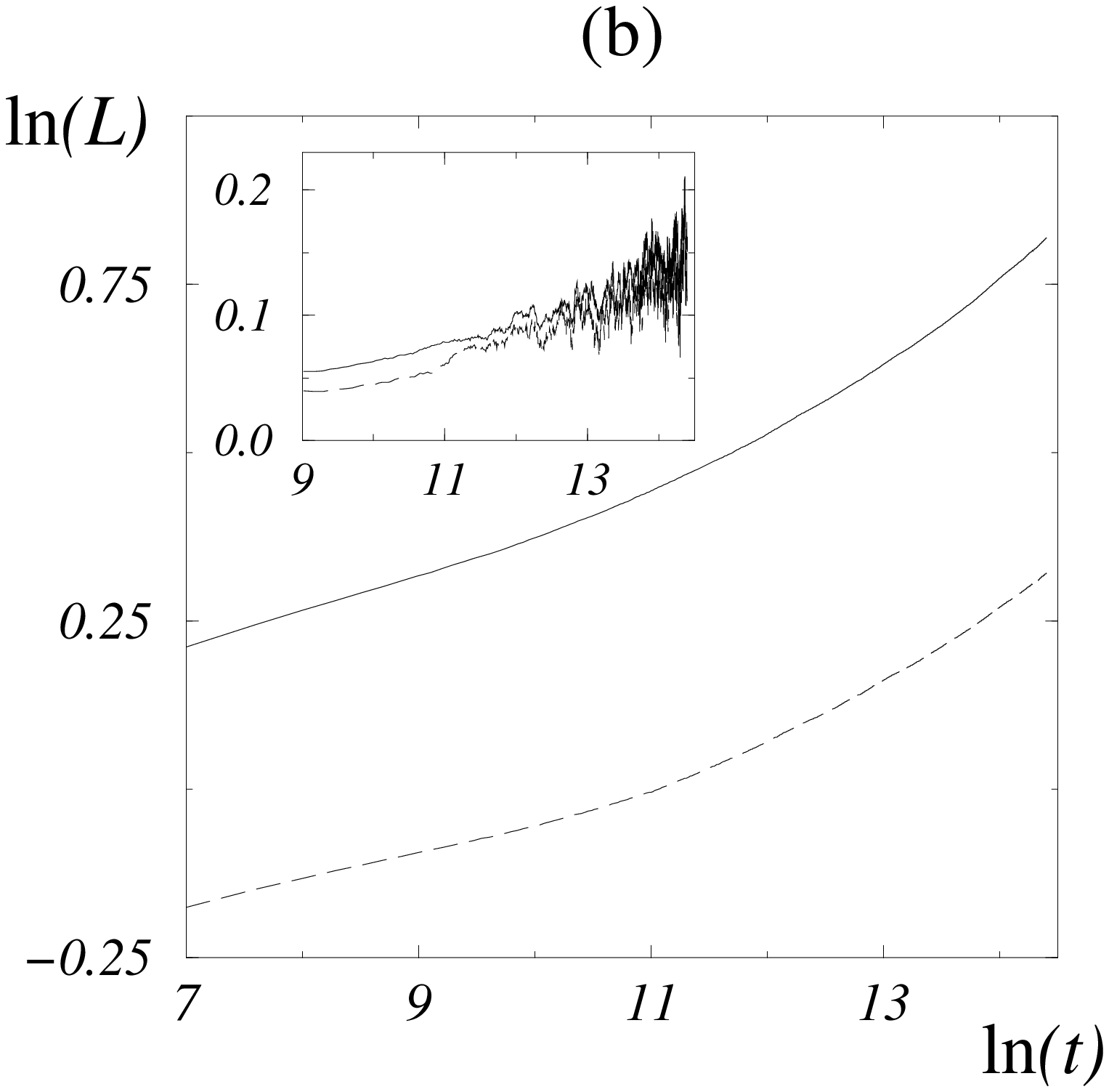}
}
\vspace{0.2cm}
\caption{Domain growth in the APS system with the local map
(\ref{mh-map}) at zero temperature ($\beta=\infty$). Lattice of size $1024^2$
with periodic boundary conditions. Solid lines: $L$ defined as the width at 
mid-height of the two-point correlation function estimated along the $x$-axis
of the lattice. Dashed lines: same but along the $y$-axis of the lattice.
(a) $L$ vs $t^{1/3}$; insert: local slopes.
(b) $\ln(L)$ vs $\ln(t)$; insert: local exponent.
}
\label{f1}
\end{figure}

We have performed numerical simulations of the APS system at zero
temperature with the
piecewise linear local map used in \cite{CMLPER}:
\begin{equation}
f(x) = \left\{
\begin{array}{lll}
\mu X & {\rm if} & X \in [-1/3,1/3] \\
2\mu/3 - \mu X & {\rm if} & X \in [1/3,1] \\
-2\mu/3 - \mu X & {\rm if} & X \in [-1,-1/3]
\end{array} \right.
\label{mh-map}
\end{equation}
with $\mu=1.9$.
Coarsening occurs, with, again, a strong anisotropy due to the mode of 
update. The growth of $L$, defined as the width at mid-height of the 
two-point autocorrelation function, is slow at short times, but then
reaches the expected $t^{1/3}$ behavior (Fig.~\ref{f1}a), 
contrary to the claims of APS. The short-time behavior may be mistaken
for anomalously slow algebraic growth (with an exponent close to the 
value $1/z=0.07$ reported by APS) when logarithmic scales 
are used, but a closer inspection shows a systematic
increase of the local exponent  (Fig.~\ref{f1}b).
As a matter of fact, the system is so anisotropic that domains elongate in time
(Fig.~\ref{f1}a). This is due to our choice of updating $x$-wise pairs
before $y$-wise pairs. Alternating this order would presumably suppress this 
effect.

Runs of the same system at finite temperatures indicate that domain
growth is faster and that the crossover to the $z=3$ behavior occurs
sooner. 
We are confident that similar results hold for the complicated
local map mostly used by APS. 
As a conclusion, Fig.~3 of \cite{APS} has to be replaced by the
variation of the {\it prefactor} of the $L\propto t^{1/3}$ law,
similarly to the final conclusions of \cite{DDAP99} for the
non-conserved order-parameter case.

\subsection{Persistence scaling is hard to measure}

The persistence probability $p(t)={\rm
Prob}\{\sigma_i(t')=\sigma_i(t_0), \; \forall t' \! \in[t_0,t]\}$ 
is usually observed to decay
algebraically with time ($p \sim (t/t_0)^{-\theta}$)
in systems with algebraic growth laws.
But persistence scaling for conserved order parameter systems is 
notoriously difficult to observe \cite{CS}. An additional difficulty lies
in the fact that the available models only show coarsening at finite
temperatures, so that one has to resort to block-scaling of
persistence. This is also the case of APS systems, {\it even at
zero temperature}, since the chaotic fluctuations of the ``couplings'' amount
 to a finite temperature. This, by the way, is the reason why
APS systems coarsen in this case.

Needless to say,
the estimates of persistence exponents presented in \cite{APS} are then
highly unreliable, if only because of the slow crossover for
 the growth law of $L$.  
Ideally, since persistence is a complex quantity involving all times
since the reference time $t_0$, one should in principle choose
$t_0$ in the asymptotic scaling regime and simulate the system 
up to times $t\gg t_0$. Given the typical values of the crossover
times (Fig.~\ref{f1}) this is hardly possible. Another
difficulty for the APS systems is the possible influence of their
strong anisotropy on the persistence exponent $\theta$ \cite{RUT}.

Rather than trying to measure properly
persistence scaling in APS systems, a possible but difficult task,
we now turn ourselves to truly deterministic models, i.e. regular
coupled map lattices, which are devoid of the drawbacks underlined
above for APS systems.

\section{Well-behaved deterministic models}

\subsection{Oono-Puri style CMLs}

Deterministic models for phase ordering of conserved systems were
introduced by Oono and Puri \cite{OONO-PURI}. Dynamical scaling 
was observed to hold for these CMLs, with a growth
law compatible with the expected $z=3$. We now present similar models
which, in addition, can be constructed for any local map.

A usual CML, such as that studied in \cite{CMLPER,DDAP99} for the non-conserved
order parameter case, can be written:
\begin{equation}
x_i^{t+1} = {\cal F}(x_i^t) \equiv (1-{\cal N}g) f(x_i^t) + 
g \sum_{j\sim i} f(x_j^t)
\label{cml-nc}
\end{equation}
where $g$ is the coupling strength, ${\cal N}$ is the number of neighbors in 
the choosen coupling range, and the sum is over these neighbors.

Following \cite{OONO-PURI}, a CML conserving exactly the continuous local
field can easily be constructed as:
\begin{equation}
x_i^{t+1} = {\cal F}(x_i^t) - \frac{1}{\cal N} \sum_{j \sim i} \left(
{\cal F}(x_j^t) - x_j^t \right)
\label{cml-co}
\end{equation}
where the last term corresponds to the extra Laplacian in the Cahn-Hilliard
equation. Conservation of the continuous field $x$ is obvious. On the other
hand, strictly speaking, the discrete field $\sigma={\rm sgn}(x)$ is not
exactly conserved and fluctuates slightly, 
because the last term in (\ref{cml-co}) may change the sign
of sites situated in domain walls. 

The synchronous mode
of update prevents excessive (i.e. other than lattice-derivated) anisotropy.
The above structure insures that ``true'' zero-temperature regimes are observed
if the local map possesses two disjoint attractors. Changing the nature of these
attractors (fixed points, limit cycles, chaotic sets), 
one can study competition between phases of different nature.

\subsection{An example}

We now present results obtained on a particular case of the models 
defined above. More comprehensive results will be reported elsewhere
\cite{TBP}.
For simplicity reasons, we again choose the map given by Eq.~(\ref{mh-map}).
For ``extra smoothness'', the Moore neighborhood on the square lattice
(${\cal N}=8$ neighbors of equal weight) was used.

\begin{figure}
\centerline{
\epsfxsize=4.1truecm
\epsffile{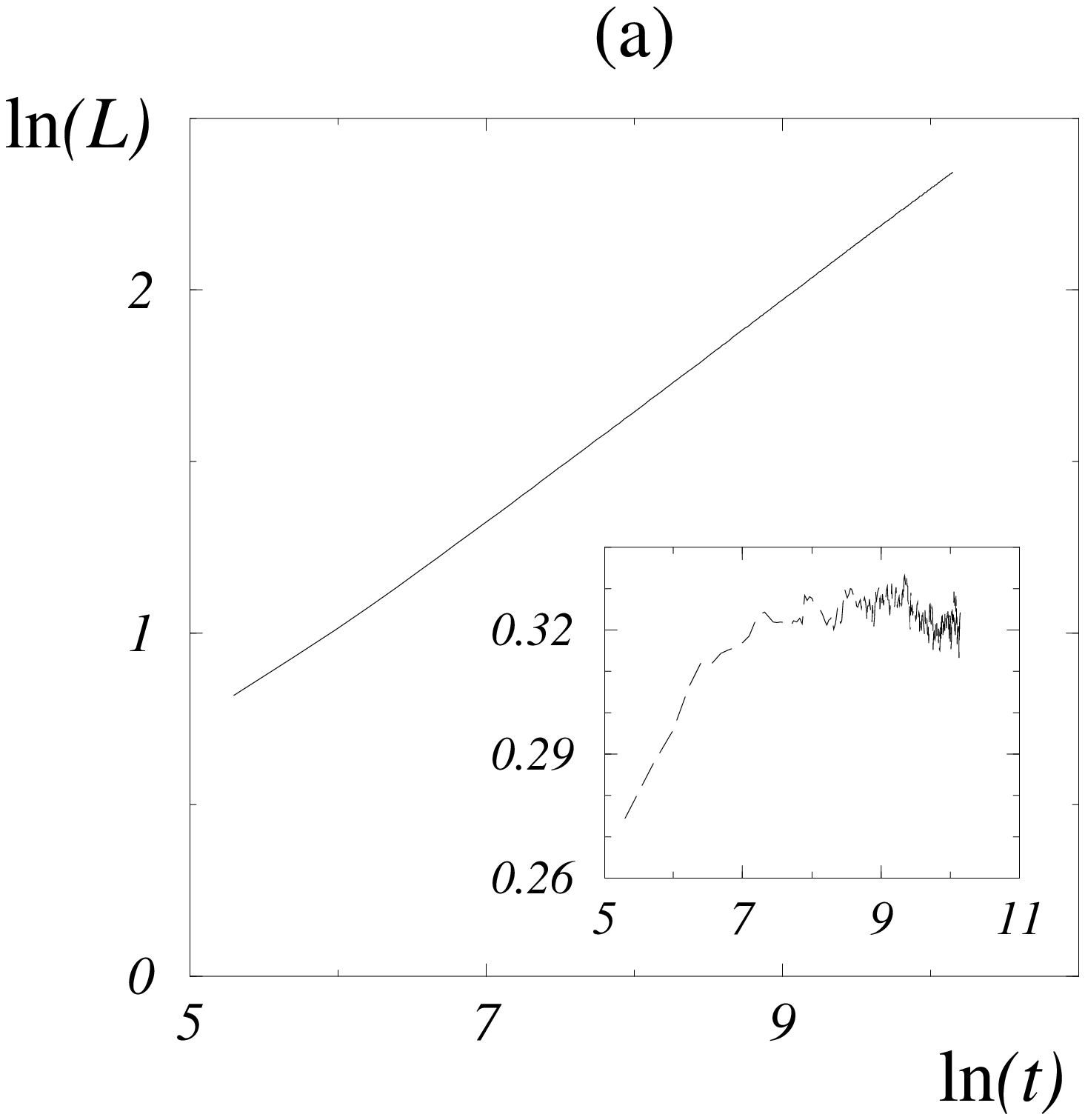}
\epsfxsize=4.3truecm
\epsffile{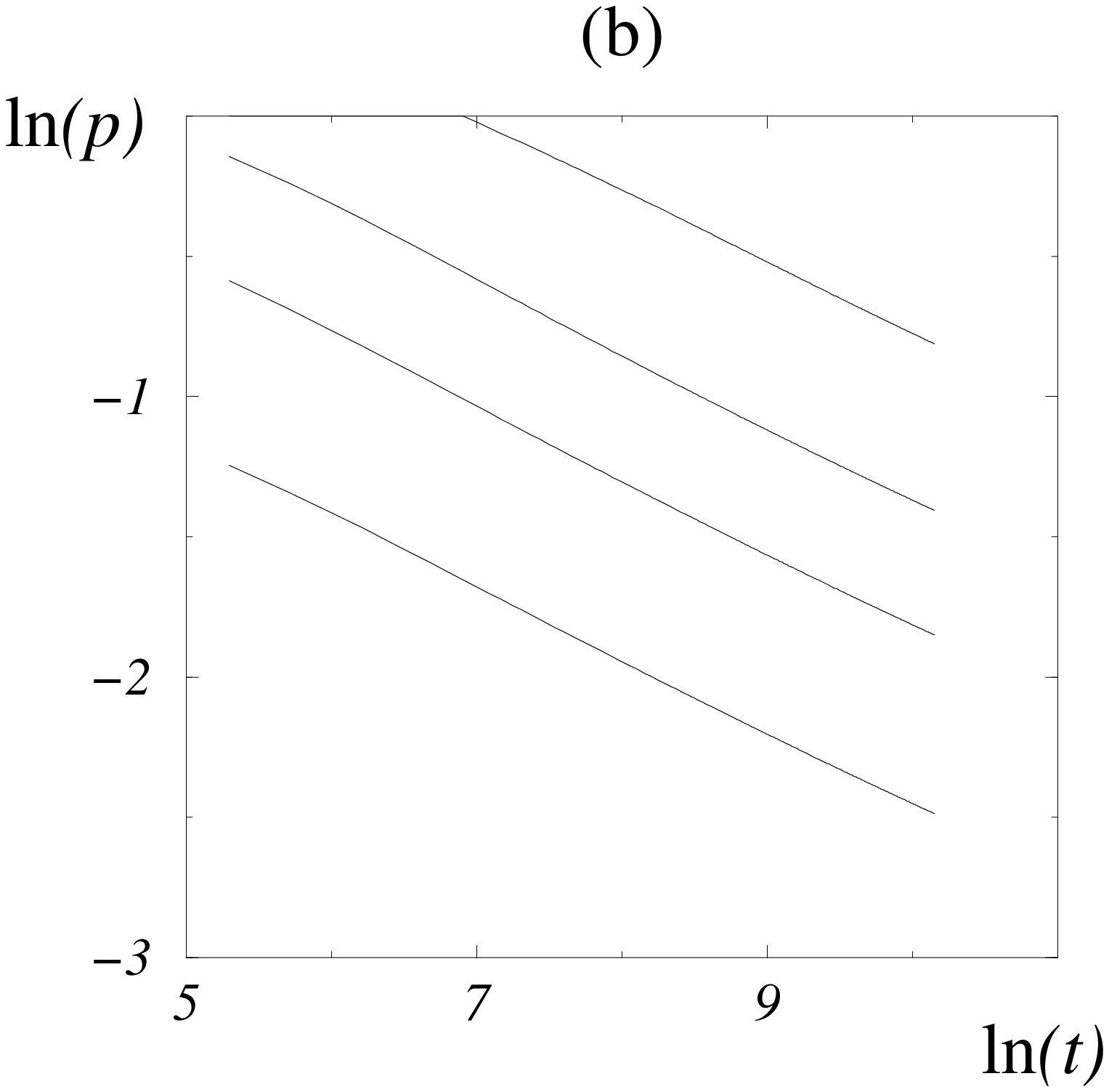}
}
\vspace{0.2cm}
\caption{Phase ordering and persistence scaling in a Oono-Puri style CML.
Lattice of size $8192^2$ with periodic boundary conditions. 
(a): $\log(L)$ vs $\log(t)$; insert: local exponent
(The slightly smaller values of $1/z$ recorded can be shown 
to be due to space discretisation effects.)
(b): $\log(p)$ vs $\log (t)$ for various initial times $t_0$. From bottom
to top: $t_0=0$, 10, $10^2$, and $10^3$.}
\label{f2}
\end{figure}

As in \cite{CMLPER,DDAP99}, domain walls are strictly pinned for small $g$.
For too strong coupling, on the other hand, antiferromagnetic-like phases
appear. There is, however, an intermediate range of $g$ values for which
domain growth proceeds forever between two weakly-chaotic phases.
The expected $z=3$ law is then easily observed even at short times 
and in log-log scales (Fig.~\ref{f2}a).

The above CMLs reveal their strongest advantage when
persistence scaling is  considered. 
As already noticed, they show normal coarsening
with true zero temperature. 
This allows to avoid studying block persistence scaling,
a somewhat tedious task. Figures~\ref{f2}b shows persistence decay for
different reference times $t_0$. Nice scaling is easily
observed. This constitutes, to our knowledge, the first clean evidence
of algebraic decay of persistence in a two-dimensional conserved order
parameter system. Our results give an exponent $\theta \simeq 0.25(2)$,
i.e. a value  larger than that observed for the non-conserved order parameter
case (for which $\theta\simeq 0.20 - 0.22$ \cite{CS,BCDL}).

The above CMLs constitute an excellent base for a reliable 
study of persistence scaling in conserved order parameter systems.
Ongoing work is probing the degree of universality of 
the persistence exponent measured
above, alongside a similar study for the non-conserved order 
parameter case for which this issue is still unresolved \cite{CS,BCDL}.

\section{Conclusion}

In this Comment, we showed how APS were misled in their interpretation of 
simulation data and that their conclusions about possible non-trivial
values of  the dynamical exponent $z$ 
in chaotic systems with conserved order parameter dynamics do not hold.
To a large extent, these systems can be seen as too close to the 
Ising model with Kawasaki dynamics, which is well-known to be difficult
to study numerically (although $z=3$ scaling is now well documented 
\cite{KAWASAKI}).

We introduced a class of CMLs which are devoid of all the problems encountered
in APS systems and which show normal scaling already at early times.
Moreover, these systems also present nice scaling behavior for the 
persistence probability, whereas similar investigations in APS systems
are riddled with problems.
We showed unambiguously that two-dimensional systems with conserved 
order parameter domain growth show an algebraic decay of persistence.
Future work will try to assess the universality of both the so-called 
Fisher-Huse exponent  and the persistence exponent $\theta$ in such systems.

\end{multicols}

\end{document}